\documentclass[conference]{IEEEtran}
\usepackage{cite}
\usepackage{amsmath,amssymb,amsfonts}
\usepackage{graphicx}
\usepackage{textcomp}
\usepackage{xcolor}
\usepackage{booktabs}
\usepackage{xurl}
\usepackage{soul}
\usepackage{color}
\begin{document}

\title{What Malware Actually Does: An Empirical Analysis of Post-Infection Behavior Across Malware Families}

\author{\IEEEauthorblockN{Emmanuela Andam\textsuperscript{1}, Emanuel S. Grant\textsuperscript{2}, and Naima Kaabouch\textsuperscript{1,2}}
\IEEEauthorblockA{\textsuperscript{1}Artificial Intelligence Research (AIR) Center, University of North Dakota, Grand Forks, ND, USA\\
\textsuperscript{2}School of Electrical Engineering and Computer Science, University of North Dakota, Grand Forks, ND, USA}}

\maketitle

\begin{center}
\footnotesize
\textcopyright~2026 IEEE. Personal use of this material is permitted. Permission from IEEE must be obtained for all other uses, in any current or future media, including reprinting/republishing this material for advertising or promotional purposes, creating new collective works, for resale or redistribution to servers or lists, or reuse of any copyrighted component of this work in other works.
\end{center}

\begin{abstract}
Most malware detection research focuses on identifying
whether a file is malicious before it executes. Far less
attention goes to what happens after infection. This paper
presents an empirical analysis of post-infection malware
behavior across 37 families, grounded in dynamic sandbox
execution using CAPEv2. Behavioral traces are modeled as
heterogeneous graphs with five node types and seven typed
edge relationships, preserving relational structure that flat
feature vectors discard. A voting ensemble trained on
graph-derived structural features achieves 94.44\% accuracy
and MCC of 0.863, outperforming a flat behavioral baseline
by 7.4 percentage points in MCC. A multiclass classifier
distinguishes 24 malware families, and a per-family
behavioral fingerprint analysis characterizes post-infection
activity across registry, network, process, and API
dimensions. An LLM explanation layer generates
analyst-readable reports with MITRE ATT\&CK mappings
grounded in observed behavioral evidence. The full pipeline
is deployed as HeteroScan at
\url{https://ama-andam.github.io/heteroscan}.
\end{abstract}

\section{Introduction}
\label{sec:intro}
 
The number of Internet of Things devices keeps growing. By
2030, connected devices are expected to exceed 32
billion~\cite{darwish_deep_2025}, including not only
embedded sensors and routers but also the Windows-based
workstations, healthcare systems, and enterprise endpoints
that form the operational backbone of critical
infrastructure. Attackers have followed that growth closely. This expansion includes not only embedded sensors and routers but also the Windows-based workstations, healthcare systems, and enterprise endpoints that form the operational backbone of critical infrastructure. In the first half of 2019, researchers recorded over 100 million IoT-targeted incidents, seven times the count from the same period in 2018~\cite{mukhtar_iot_2023}. At least 64 distinct malware families now target these devices~\cite{gulatas_malware_2023}, and the malware itself has become more capable over time. The number of device-specific exploits doubled annually after 2017~\cite{al_alsadi_no_2022}, and families that once ran simple DDoS operations have expanded into ransomware, cryptominers, and persistent spyware~\cite{victor_iot_2023}. Malware no longer targets one platform or one sector. In 2025, IoT attacks increased by 400\%, Linux infections rose by 35\%, and cryptojacking incidents surged by 659\%~\cite{ferdous_survey_2025}. Ransomware spread across desktops, mobile devices, cloud systems, and industrial infrastructure in the same year. The threat is no longer confined; it is everywhere, and it is adaptive.
 
Despite this growth, the defenses most widely deployed were not designed for it. Signature-based detection, the approach built into most endpoint tools, works by matching known malware patterns against a database of prior samples. Against catalogued threats, it is fast and accurate. Against new variants and obfuscated code, it falls apart. Empirical evaluations place zero-day detection rates as low as 8.2\% for signature-only systems~\cite{holm_signature_2014}, and a few bytes of mutation can bypass byte-pattern matching entirely~\cite{sihwail_survey_2018}. Machine learning models trained on static features face the same problem from a different angle: they generalize from historical distributions that may not reflect the variants attackers release today~\cite{guo_survey_2023}.

\begin{figure}[t]
\centering
\includegraphics[width=\columnwidth]{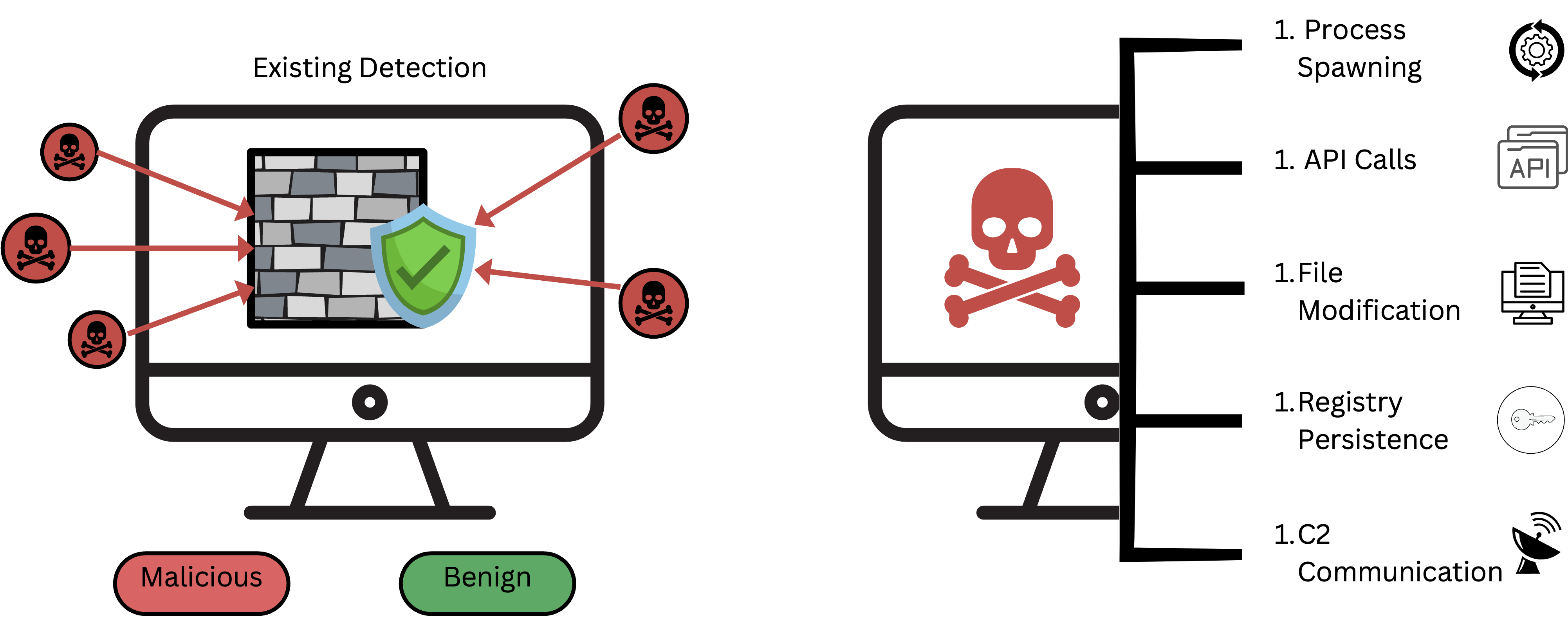}
\caption{Existing detection approaches classify executables as malicious or benign at the perimeter but do not analyze post-infection behavior. This work addresses the gap by empirically characterizing what malware actively does inside a compromised system.}
\label{fig:gap}
\end{figure}
\begin{figure*}[t]
\centering
\includegraphics[width=\textwidth]{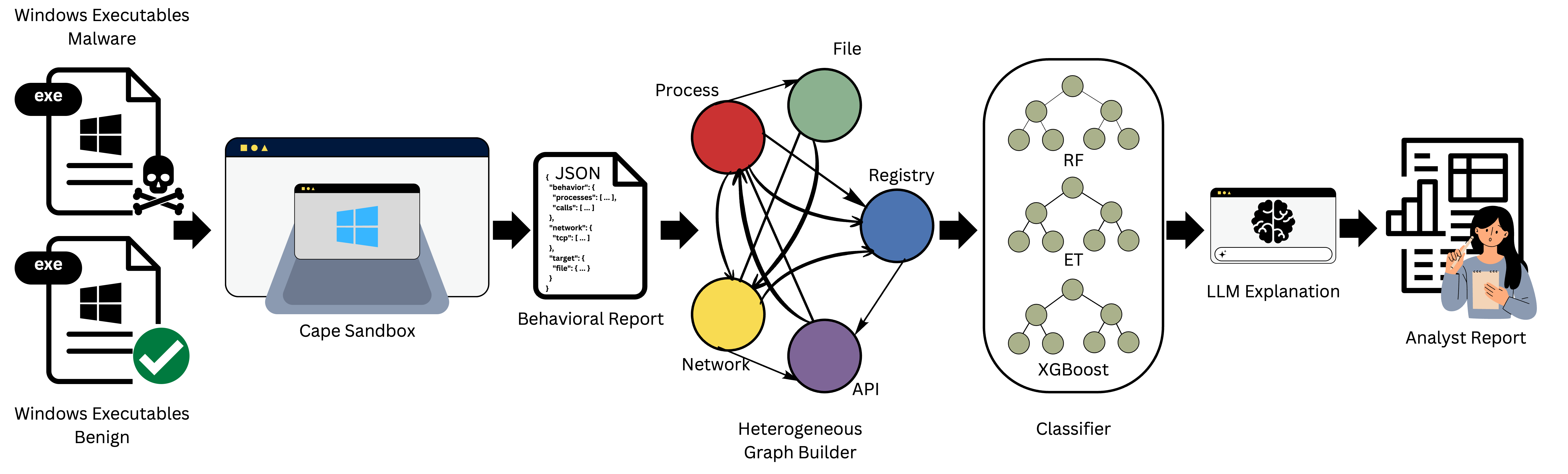}
\caption{HeteroScan pipeline: Windows executables (malicious and benign) are submitted to the CAPE sandbox, behavioral reports are parsed into heterogeneous graphs, structural features are extracted for ensemble classification, and an LLM generates analyst-readable behavioral reports.}
\label{fig:pipeline}
\end{figure*}
Dynamic sandbox analysis gets further than static methods. By executing a suspicious sample in isolation and recording what it does at runtime, sandboxes can surface process creation, file and registry changes, API call sequences, and command-and-control traffic that static inspection cannot observe~\cite{mahmoud_redefining_2024,jain_dynamic_2025}. One large-sample study found suspicious runtime activity in 95\% of analyzed samples~\cite{khalda_analisis_2025}. Even so, most research that uses dynamic analysis asks only one question: is this sample malicious? The behavioral telemetry collected during execution reveals much more than a binary verdict. It shows what the malware is actively doing to a compromised system, and that information rarely becomes the main object of study.
 
That gap matters in practice. Knowing a device is infected is only the starting point for incident response. Analysts need to understand what happened after infection: which processes were spawned, which files were created or encrypted, which registry keys were written for persistence, and which external endpoints were contacted. Most behavioral analysis tools answer this by producing flat API call logs or aggregate feature vectors, losing the relational structure between entities. The fact that a specific process wrote to a specific registry key while simultaneously connecting to a non-standard external port carries meaning that a list of API frequencies does not capture.

Prior work on behavioral malware analysis has largely relied on two approaches: flat feature vectors extracted from sandbox logs, and homogeneous API call graphs where all behavioral signals collapse into a single node type. Singh et al.~\cite{singh_detection_2020} and Jain et al.~\cite{jain_dynamic_2025} show that sandbox-derived behavioral features fed into tree-based classifiers achieve strong detection rates, but flattening execution telemetry into vectors loses the relational structure between entities. Graph-based approaches address part of this by modeling API call sequences as directed graphs~\cite{elhadi_improving_2013,pektas_deep_2020,li_dmalnet_2022}, achieving up to 98.86\% accuracy, but treating all entities as the same node type discards the semantic distinctions between processes, files, registry keys, and network endpoints. Heterogeneous graph neural networks have shown promise in network intrusion detection~\cite{farrukh_xg-nid_2025,jiang_survey_2026}, where modeling multiple entity types improves detection over homogeneous baselines, yet this formulation remains largely absent from post-infection malware behavioral analysis. On the explanation side, recent work shows that LLMs can generate human-readable behavioral summaries from sandbox reports~\cite{saha_malaware_2025} and static code~\cite{fujii_feasibility_2024}, reducing analyst workload during incident response. Our work combines these directions: we model post-infection behavior as a heterogeneous graph across five entity types and seven relationship types, classify using an ensemble trained on graph-derived features, and generate analyst-readable explanations through an LLM layer grounded in observed behavioral evidence. This study extends our prior work on malware detection, which addressed few-shot detection of emerging variants~\cite{andam2026maml} and multi-class malware classification using structured state-space models~\cite{andam2026s4}; the present study is part of a broader investigation of malware detection and behavioral analysis~\cite{andam2026thesis}, shifting focus from detection to post-infection behavioral characterization.
 
This paper addresses that gap directly. We focus on what malware actually does once it is inside a system, analyzing samples from multiple malware families through dynamic sandbox execution. The contributions of this paper are as follows:
\begin{itemize}
    \item A heterogeneous behavioral graph construction pipeline that models post-infection behavior across multiple entity types and relationship types
    \item An empirical characterization of post-infection behavioral patterns across a range of malware families
    \item A classification framework trained and evaluated on behavioral graph features extracted from sandbox-analyzed executables
    \item An LLM-based explanation layer that produces natural language behavioral summaries and maps detected behaviors to MITRE ATT\&CK techniques
\end{itemize}
\begin{figure}[t]
\centering
\includegraphics[width=\columnwidth]{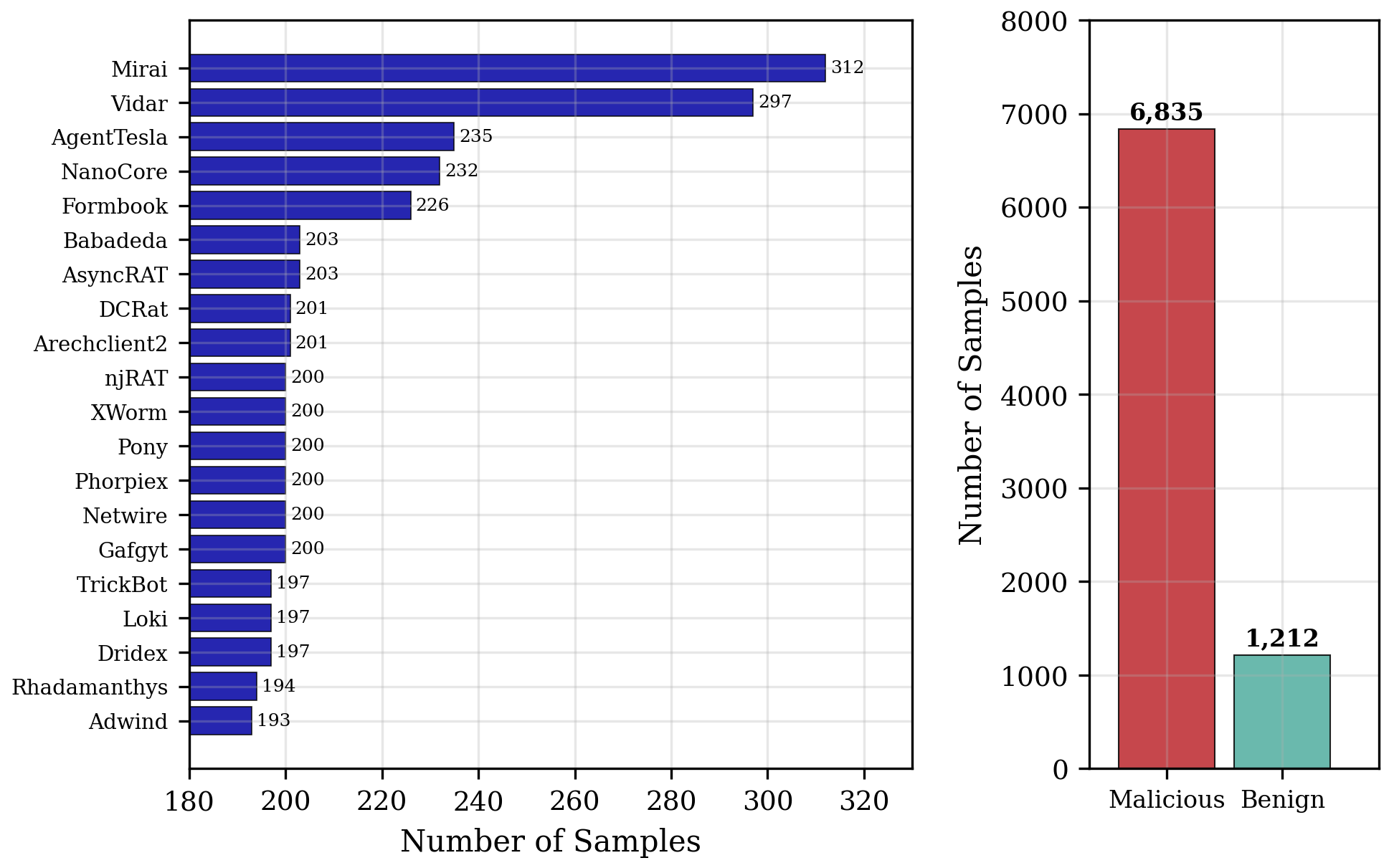}
\caption{Dataset overview: top 20 malware families by sample count (left) and overall dataset composition (right). The full dataset comprises 8,047 samples across 37 malware families and benign executables from four sources.}
\label{fig:dataset}
\end{figure}

\section{Methodology}
\label{sec:methodology}

\subsection{System Overview}

Analyzing what malware does after gaining access to a system
requires moving beyond static file inspection. Static
approaches examine a binary's code, imports, and structure
without executing it, which means they can only infer
behavior rather than observe it directly. Our system takes a
different approach: it executes each sample in a controlled
environment, records everything the sample does at runtime,
and builds a structured representation of those observations
that can be analyzed and explained.

The system consists of four components working in sequence.
The first is a dynamic analysis environment based on the CAPE
sandbox, which executes Windows executables in an isolated
virtual machine and generates structured behavioral reports
covering process activity, file system interactions, registry
modifications, API call sequences, and network
communications. The second is a graph construction pipeline
that parses each behavioral report into a heterogeneous
graph, where five types of system entities become nodes and
seven types of interactions between them become typed edges.
This representation preserves the relational structure of
post-infection behavior that flat feature vectors discard.
The third component is a classification framework that
extracts 55 structural features from each behavioral graph
and passes them to a voting ensemble of three classifiers:
Random Forest, Extra Trees, and XGBoost, to produce a
malicious or benign verdict with a decision-threshold-tuned
confidence score. The fourth component is an LLM-based
explanation layer that receives the behavioral evidence and
classifier output and generates a natural language analysis
report covering observed threat indicators and MITRE
ATT\&CK technique mappings.

The full pipeline is illustrated in Fig.~\ref{fig:pipeline}.
The system is also deployed as HeteroScan, a web-based tool
that accepts executable uploads, runs them through the full
pipeline, and returns an analyst-readable behavioral report.

\subsection{Dynamic Analysis Environment}

The dynamic analysis environment is built on
CAPEv2~\cite{noauthor_cape_nodate}, an open-source sandbox framework
designed for malware behavioral analysis. CAPE executes
submitted samples inside an isolated virtual machine and
monitors system activity at the kernel level, producing
structured JSON reports that capture the full execution
trace. We selected CAPE over alternative sandboxes such as
Cuckoo and Any.run because it provides direct access to raw
behavioral telemetry through a REST API, supports custom
virtual machine configurations, and allows automated bulk
sample submission.

The analysis virtual machine runs Windows 10 on a KVM/QEMU
hypervisor. Microsoft Defender was disabled through Windows
registry policy by setting \texttt{DisableAntiSpyware},
\texttt{DisableRealtimeMonitoring}, and
\texttt{DisableIOAVProtection} to \texttt{1}, and all
Defender-related services were disabled offline via
\texttt{virt-win-reg}. Windows Update was disabled to
prevent background network traffic from interfering with
command-and-control observations, and network access was
left enabled so that samples attempting to contact external
servers produce observable activity. Several .NET-dependent
families including NanoCore required .NET Framework 3.5,
which was installed offline via DISM. A clean snapshot was
taken after configuration and is restored before each sample
execution, ensuring a consistent environment across all
analyses.

\subsection{Dataset Collection}
The dataset consists of two classes of Windows Portable
Executable (PE) files: malicious samples collected from
MalwareBazaar and benign samples collected from four
independent sources. Ground truth labels were assigned by
source rather than by antivirus consensus, avoiding the
label noise introduced when antivirus consensus labels are
used as ground truth~\cite{joyce_motif_2023}.
\begin{table}[t]
\caption{Per-Family Graph Survival Rates (Selected Families)}
\label{tab:filtering}
\centering
\scriptsize
\begin{tabular}{lrrr}
\toprule
\textbf{Family} & \textbf{Submitted} & \textbf{Rich Graphs}
& \textbf{Rate} \\
\midrule
Vidar        & 65  & 61  & 94\% \\
Formbook     & 40  & 36  & 90\% \\
njRAT        & 54  & 47  & 87\% \\
XWorm        & 76  & 51  & 67\% \\
NanoCore     & 205 & 93  & 45\% \\
AsyncRAT     & 244 & 91  & 37\% \\
DCRat        & 263 & 73  & 28\% \\
Phorpiex     & 200 & 0   & 0\%  \\
\bottomrule
\end{tabular}
\end{table}
Malicious samples were downloaded from the MalwareBazaar API
across 37 families using the \texttt{get\_siginfo} endpoint.
Families include remote access trojans such as AsyncRAT,
XWorm, njRAT, QuasarRAT, and NanoCore; infostealers such as
RedLine, AgentTesla, Vidar, Stealc, and Formbook; loaders
such as GuLoader, PrivateLoader, and Amadey; banking trojans
such as TrickBot, Dridex, and IcedID; and IoT botnets such
as Mirai and Gafgyt, among others. Benign samples came from
the DikeDataset repository, PortableApps installers, GitHub
releases, and Windows system executables from the analysis
VM. The dataset comprises 6,835 malicious and 1,212 benign
samples totaling 8,047 executables.

After graph construction, samples with fewer than two unique
API nodes were excluded as they exhibited insufficient
behavioral activity. This removes samples that evaded the
sandbox or terminated before generating observable content.
The resulting dataset comprises 2,516 rich behavioral
graphs: 1,793 malicious and 723 benign, partitioned into
70\% training (1,760 samples), 15\% validation (378
samples), and 15\% test (378 samples) using stratified
sampling. Table~\ref{tab:filtering} summarizes per-family
survival rates for selected families.

Families with zero surviving graphs, such as Phorpiex,
terminated within seconds of execution after performing
brief environment checks. These families are excluded from
the classification experiments. Behavioral differences
between malicious and benign samples may partly reflect
collection source rather than maliciousness alone, as all
malicious samples originate from MalwareBazaar while benign
samples come from four distinct sources.

\subsection{Heterogeneous Graph Construction}

\begin{figure}[t]
\centering
\includegraphics[width=\columnwidth]{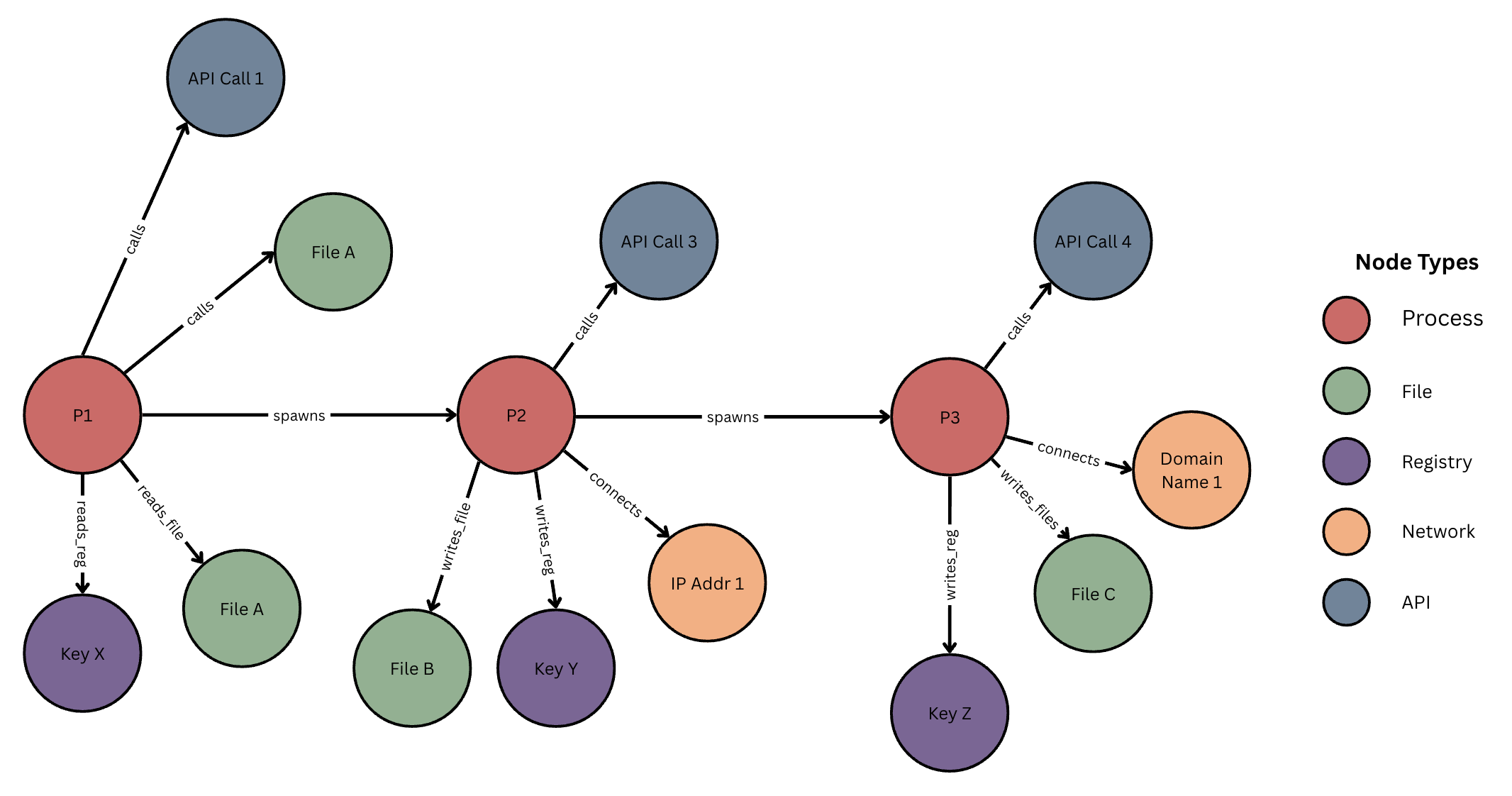}
\caption{Heterogeneous behavioral graph structure. Five node
types and seven typed edge relationships capture the
post-infection system entity interactions observed during
sandbox execution.}
\label{fig:graph}
\end{figure}

Each CAPE sandbox report is parsed into a PyTorch Geometric
\texttt{HeteroData} object. The graph captures five node
types and seven edge types as defined in
Tables~\ref{tab:nodes} and~\ref{tab:edges}, and illustrated
in Fig.~\ref{fig:graph}.

\begin{table}[t]
\caption{Node Types and Descriptions}
\label{tab:nodes}
\centering
\begin{tabular}{ll}
\toprule
\textbf{Node Type} & \textbf{Description} \\
\midrule
Process  & Executable observed spawning during execution \\
File     & File path read from or written to \\
Registry & Registry key queried or modified \\
Network  & External endpoint contacted via TCP or UDP \\
API      & Windows API function called by a process \\
\bottomrule
\end{tabular}
\end{table}

\begin{table}[b]
\caption{Edge Types and Semantic Meaning}
\label{tab:edges}
\centering
\begin{tabular}{lll}
\toprule
\textbf{Edge Type} & \textbf{Source$\to$Target} &
\textbf{Condition} \\
\midrule
spawns       & Process$\to$Process  & Parent PID $\neq$ child PID \\
calls        & Process$\to$API      & Any API call \\
reads\_file  & Process$\to$File     & API contains Read/Open/Query \\
writes\_file & Process$\to$File     & All other file API calls \\
reads\_reg   & Process$\to$Registry & API contains Open/Query \\
writes\_reg  & Process$\to$Registry & All other registry API calls \\
connects     & Process$\to$Network  & Network API calls and TCP/UDP \\
\bottomrule
\end{tabular}
\end{table}

Node feature matrices are initialized as ones vectors of
dimension one per node, as no semantic per-node features are
available from the sandbox reports. Classification therefore
relies entirely on graph-level structural statistics rather
than on learned node feature representations; the 30
node-type statistical features in Table~\ref{tab:features}
summarize these uninformative node vectors and serve
primarily as a compact encoding of node count distributions
across types. Graphs with fewer than two unique API nodes
are excluded from classifier training, yielding 2,516 graphs
with sufficient behavioral content comprising 1,793
malicious and 723 benign samples.

\subsection{Feature Extraction and Classification}

A 55-dimensional feature vector is extracted from each
behavioral graph as summarized in Table~\ref{tab:features}.
Features capture the behavioral signal encoded in the graph
topology through node counts, pairwise ratios, log
transforms, per-node-type statistical summaries, and edge
counts per relationship type. All features are standardized
using z-score normalization fit on the training set.

\begin{table}[t]
\caption{Feature Categories (55 Total)}
\label{tab:features}
\centering
\begin{tabular}{lcc}
\toprule
\textbf{Category} & \textbf{Count} & \textbf{Description} \\
\midrule
Raw node counts  & 5  & Count per node type \\
Total node count & 1  & Sum of all node counts \\
Pairwise ratios  & 7  & API/proc, reg/proc, net/proc, etc. \\
Log transforms   & 5  & $\log(1 + n)$ per node type \\
Node statistics  & 30 & Mean, std, max, min, sum, active \\
                 &    & fraction per node type (6$\times$5) \\
Edge counts      & 7  & Count per edge type \\
\bottomrule
\end{tabular}
\end{table}

Three classifiers were trained as summarized in
Table~\ref{tab:classifiers}. A voting ensemble was formed
by averaging the predicted class probabilities. The decision
threshold was tuned on the validation set by maximizing MCC.
To evaluate whether graph-derived features outperform
simpler representations, we compare the ensemble against a
flat five-feature baseline comprising raw node counts only,
using identical architectures and the same test split.
Performance is evaluated using accuracy, F1-score,
precision, recall, and MCC, with MCC preferred for its
robustness to class imbalance. For multiclass classification,
macro and weighted F1-scores are reported.

\begin{table}[b]
\caption{Classifier Hyperparameters}
\label{tab:classifiers}
\centering
\begin{tabular}{ll}
\toprule
\textbf{Classifier} & \textbf{Key Hyperparameters} \\
\midrule
Random Forest   & n\_estimators=500, max\_features=sqrt, \\
                & class\_weight=balanced \\
Extra Trees     & n\_estimators=500, max\_features=sqrt, \\
                & class\_weight=balanced \\
XGBoost         & n\_estimators=500, max\_depth=6, \\
                & learning\_rate=0.05, subsample=0.8, \\
                & scale\_pos\_weight=benign/malicious ratio \\
Voting Ensemble & Average RF + ET + XGBoost probabilities \\
\bottomrule
\end{tabular}
\end{table}

\subsection{LLM Explanation Layer}

Following classification, a behavioral summary is assembled
from the graph metadata including spawned process names,
unique API functions called, file paths written, registry
keys accessed, and network endpoints contacted. This summary
is passed to Claude Sonnet 4.6 via the Anthropic API
together with the classifier verdict and confidence score.
The model produces a structured report covering: a verdict
statement with confidence, a two to three sentence
behavioral summary, the three most suspicious behavioral
indicators, a malware family estimate, and MITRE ATT\&CK
technique identifiers. The prompt explicitly restricts the
model to the observed behavioral evidence, reducing
hallucinated claims about behavior not observed during
execution~\cite{conceicao_evaluation_2025}. The explanation layer
does not perform classification; it generates natural
language descriptions of the behavioral evidence that the
ensemble classifier acted on, keeping the detection decision
auditable at the level of graph structure while adding
analyst-readable context.
\begin{figure}[t]
\centering
\includegraphics[width=0.4\textwidth]{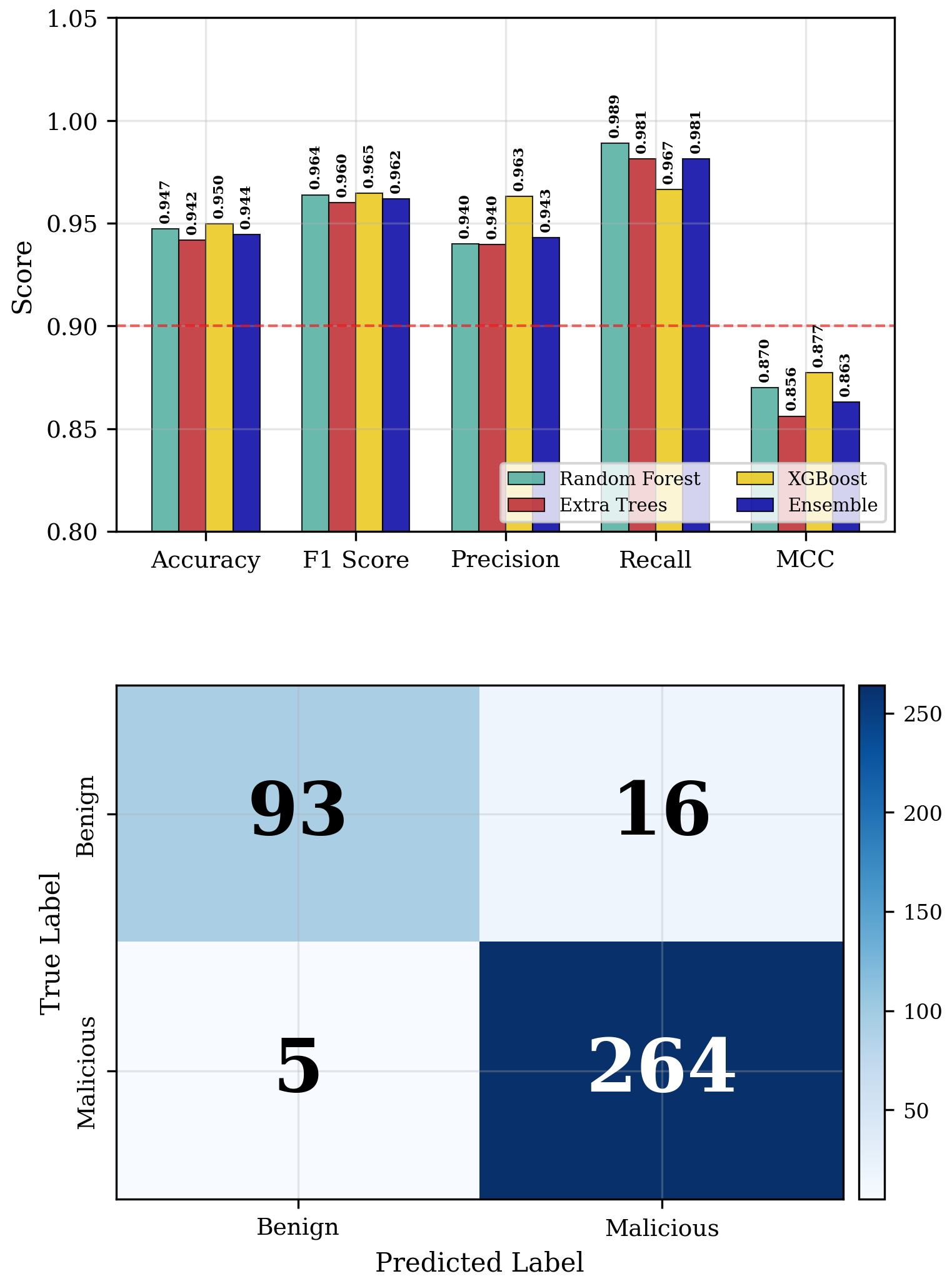}
\caption{Model performance comparison across five metrics on
the test set (left) and confusion matrix for the voting
ensemble (right). The dashed red line indicates the 90\%
accuracy threshold.}
\label{fig:comparison}
\end{figure}

\section{Results and Discussion}
\label{sec:results}

\subsection{Experimental Setup}

All experiments were conducted on a workstation with an
Intel Core i7-12700 processor and 16GB RAM running Ubuntu
22.04. The ensemble classifier was implemented using
scikit-learn 1.3 and XGBoost 3.2. The LLM explanation layer
used Claude Sonnet 4.6 via the Anthropic API.

\subsection{Binary Detection Performance}

Fig.~\ref{fig:comparison} presents the performance
comparison across all four models. Table~\ref{tab:results}
reports the full numerical results on the 378-sample test
set (269 malicious, 109 benign). The voting ensemble
achieves 94.44\% accuracy, an F1 score of 0.9617, and an
MCC of 0.8628. Among individual classifiers, XGBoost
achieves the highest precision at 96.30\% with only 10
false positives, making it the most conservative model.
Random Forest achieves the highest recall at 98.88\% with
only 3 false negatives, making it the most sensitive to
malicious activity. The ensemble balances these properties,
achieving an MCC of 0.8628. ROC analysis confirms
these results, with the ensemble achieving the highest AUC
of 0.966, followed by Random Forest (0.963), Extra Trees
(0.958), and XGBoost (0.955). The confusion matrix shows
TN=93, FP=16, FN=5, TP=264.

\begin{table}[t]
\caption{Classification Results on Test Set (378 samples:
269 malicious, 109 benign)}
\label{tab:results}
\centering
\begin{tabular}{lccccc}
\toprule
\textbf{Model} & \textbf{Acc} & \textbf{F1} & \textbf{Prec}
& \textbf{Rec} & \textbf{MCC} \\
\midrule
Random Forest & 0.9471 & 0.9638 & 0.9399 & 0.9888 & 0.8698 \\
Extra Trees   & 0.9418 & 0.9600 & 0.9395 & 0.9814 & 0.8561 \\
XGBoost       & 0.9497 & 0.9647 & 0.9630 & 0.9665 & 0.8772 \\
\textbf{Ensemble} & \textbf{0.9444} & \textbf{0.9617} &
\textbf{0.9429} & \textbf{0.9814} & \textbf{0.8628} \\
\bottomrule
\end{tabular}
\end{table}

\subsection{Graph Features vs.\ Flat Behavioral Counts}

To evaluate whether the heterogeneous graph representation
provides advantage over simpler behavioral features,
Table~\ref{tab:flat} compares the ensemble trained on the
full 55-dimensional graph-derived feature vector against an
ensemble trained on a flat five-dimensional behavioral count
vector comprising raw process, file, registry, network, and
API node counts. Both ensembles use identical architectures,
hyperparameters, and train/test splits.

\begin{table}[t]
\caption{Graph-Derived Features vs.\ Flat Behavioral Counts}
\label{tab:flat}
\centering
\begin{tabular}{lcccc}
\toprule
\textbf{Representation} & \textbf{Features} & \textbf{Acc}
& \textbf{F1} & \textbf{MCC} \\
\midrule
Flat behavioral counts & 5  & 0.9127 & 0.9399 & 0.7828 \\
Graph-derived features & 55 & 0.9418 & 0.9596 & 0.8565 \\
\midrule
\textbf{Improvement} & +50 & +0.029 & +0.020 &
\textbf{+0.074} \\
\bottomrule
\end{tabular}
\end{table}

The graph-derived feature vector outperforms the flat count
baseline across all three metrics, with the most pronounced
gap in MCC (+7.4 percentage points). This confirms that
typed behavioral interaction counts, particularly edge
counts per relationship type, carry discriminative signal
beyond what raw entity counts provide.

\subsection{Feature Importance Analysis}
\label{sec:importance}

\begin{figure}[t]
\centering
\includegraphics[width=\columnwidth]{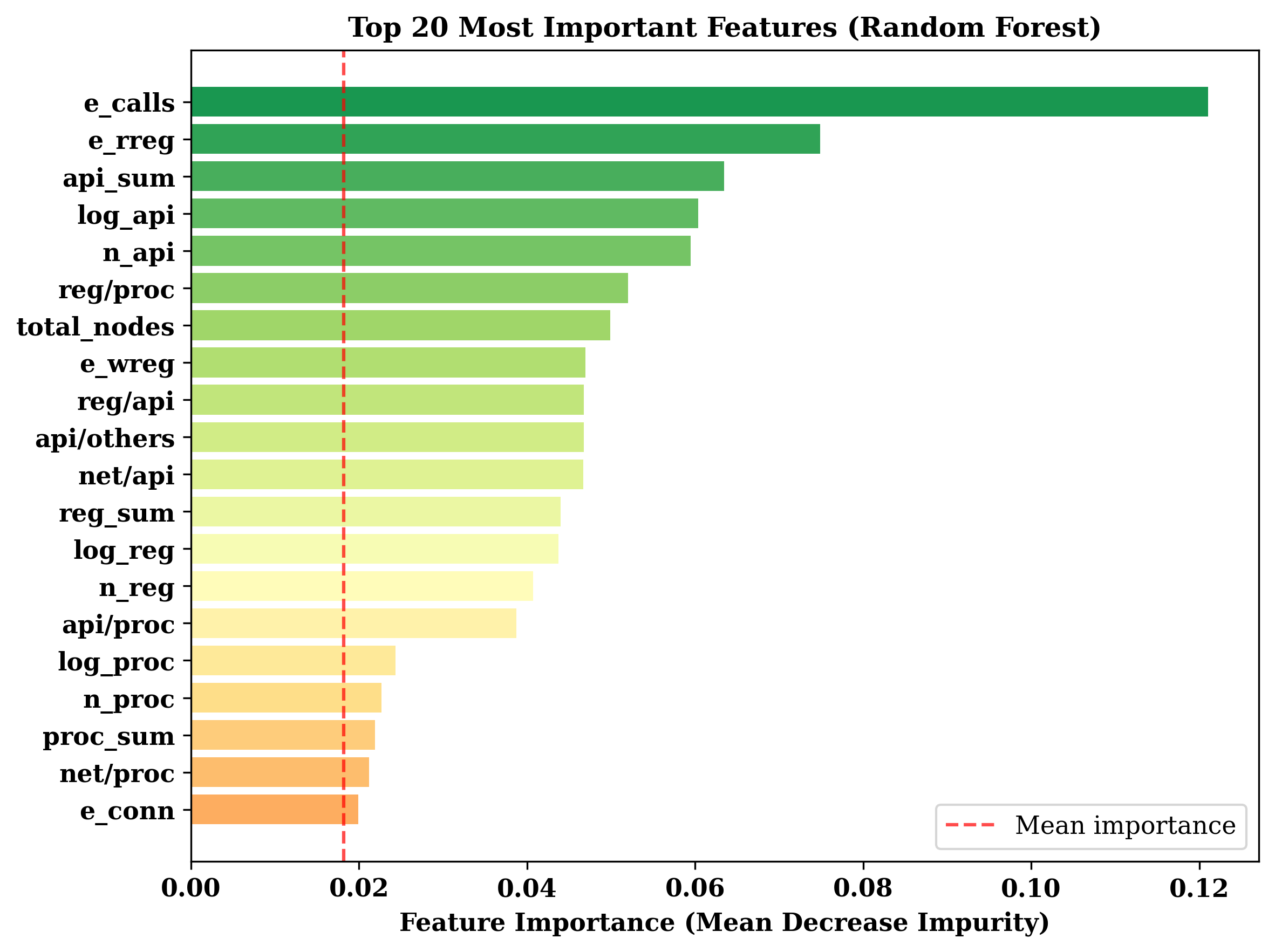}
\caption{Top 20 most important features ranked by mean
decrease impurity from the Random Forest classifier. Edge
count features, particularly \texttt{e\_calls} and
\texttt{e\_reads\_reg}, dominate the ranking.}
\label{fig:importance}
\end{figure}

Fig.~\ref{fig:importance} shows the top 20 most important
features from the Random Forest classifier. The most
discriminative feature by a substantial margin is
\texttt{e\_calls}, the count of calls edges, representing
the total number of process-to-API interactions. The second
most important is \texttt{e\_reads\_reg}, the count of
registry read edges. Edge-based features dominate the top
rankings, indicating that typed behavioral interaction
counts are more discriminative than raw entity counts in
this dataset. Registry-related features appear repeatedly,
reflecting a consistent behavioral pattern: malicious
samples systematically enumerate registry keys as part of
environment fingerprinting and persistence establishment.

\subsection{Comparison with GNN Baseline}

The heterogeneous GNN was evaluated on the original
1,554-graph filtered dataset prior to dataset expansion.
Table~\ref{tab:gnn} summarizes the comparison. The GNN
achieved 76.25\% accuracy and MCC of 0.382, substantially
below the ensemble. The FP/FN counts of 141/30 reflect
evaluation on the full 1,554-graph dataset rather than the
378-sample test split used for the ensemble, and are
therefore not directly comparable on a per-sample basis.
The underperformance stems from uninformative node features:
all node feature matrices are initialized as placeholder
ones vectors, providing no semantic signal for message
passing. The GNN result should therefore be treated as
preliminary; a fairer comparison would require semantic node
features such as API category embeddings or registry key
type labels.

\begin{table}[t]
\caption{Ensemble vs.\ GNN Baseline}
\label{tab:gnn}
\centering
\begin{tabular}{lcccc}
\toprule
\textbf{Model} & \textbf{Acc} & \textbf{F1} & \textbf{MCC}
& \textbf{FP/FN} \\
\midrule
Heterogeneous GNN\textsuperscript{$\dagger$} & 0.763 & 0.846
& 0.382 & 141 / 30 \\
\textbf{Ensemble} & \textbf{0.944} & \textbf{0.962} &
\textbf{0.863} & \textbf{16 / 5} \\
\bottomrule
\end{tabular}
\end{table}
\textsuperscript{$\dagger$}GNN evaluated on the full
1,554-graph dataset prior to expansion; FP/FN counts are
not directly comparable to ensemble test split results.

\subsection{Multiclass Family Classification}

To move beyond binary detection and address the paper's
central question of what malware does across families, we
train the voting ensemble for multiclass malware family
classification. Training uses only malicious samples from
families with at least 10 behavioral graphs, yielding 1,774
samples across 24 families with a 70/15/15 split.

\begin{table}[t]
\caption{Multiclass Family Classification Results}
\label{tab:multiclass}
\centering
\begin{tabular}{lc}
\toprule
\textbf{Metric} & \textbf{Value} \\
\midrule
Accuracy        & 0.6891 \\
Macro F1        & 0.5818 \\
Weighted F1     & 0.6773 \\
Families        & 24 \\
Test samples    & 267 \\
\bottomrule
\end{tabular}
\end{table}

The ensemble achieves 68.91\% accuracy and macro F1 of
0.5818 across 24 families. NanoCore achieves the highest
F1 of 0.974, reflecting its distinctive .NET Framework
registry enumeration. Babadeda (0.857), BlackMatter (0.857),
and Vidar (0.900) also perform strongly. Families with fewer
test samples such as IcedID and Netwire score near zero,
reflecting insufficient representation. The lower macro F1
relative to binary detection is expected given the
difficulty of distinguishing 24 families using aggregate
structural features alone.

\subsection{Behavioral Fingerprint Analysis}

Table~\ref{tab:heatmap} presents median behavioral
statistics per family, providing an empirical
characterization of what each malware family does
post-infection.

\begin{table}[t]
\caption{Median Behavioral Statistics Per Family}
\label{tab:heatmap}
\centering
\scriptsize
\begin{tabular}{lrrrrr}
\toprule
\textbf{Family} & \textbf{Proc} & \textbf{File} &
\textbf{Reg} & \textbf{Net} & \textbf{API} \\
\midrule
BlackMatter    & 5 & 1 & 5036 & 1 & 118 \\
QuasarRAT      & 5 & 1 & 1190 & 2 & 147 \\
WSHRAT         & 3 & 1 & 1136 & 2 & 98  \\
DCRat          & 5 & 1 & 1025 & 1 & 137 \\
NanoCore       & 5 & 1 & 734  & 3 & 143 \\
Babadeda       & 5 & 1 & 814  & 2 & 102 \\
Amadey         & 2 & 1 & 513  & 2 & 101 \\
Limerat        & 1 & 1 & 495  & 1 & 98  \\
SnakeKeylogger & 2 & 1 & 431  & 1 & 117 \\
Netwire        & 4 & 1 & 422  & 1 & 122 \\
AsyncRAT       & 3 & 1 & 342  & 1 & 104 \\
Arechclient2   & 4 & 1 & 338  & 2 & 92  \\
Formbook       & 3 & 1 & 302  & 2 & 106 \\
njRAT          & 3 & 1 & 286  & 1 & 106 \\
PrivateLoader  & 1 & 1 & 194  & 2 & 96  \\
AgentTesla     & 3 & 1 & 200  & 2 & 76  \\
Azorult        & 1 & 1 & 140  & 2 & 82  \\
Vidar          & 1 & 1 & 103  & 2 & 70  \\
Stealc         & 1 & 1 & 98   & 4 & 80  \\
Bumblebee      & 3 & 1 & 85   & 2 & 68  \\
IcedID         & 3 & 1 & 85   & 2 & 68  \\
Bandook        & 1 & 1 & 64   & 3 & 80  \\
GuLoader       & 1 & 1 & 58   & 2 & 81  \\
XWorm          & 1 & 1 & 14   & 1 & 36  \\
\bottomrule
\end{tabular}
\end{table}

\begin{figure*}[h]
\centering
\includegraphics[width=\textwidth]{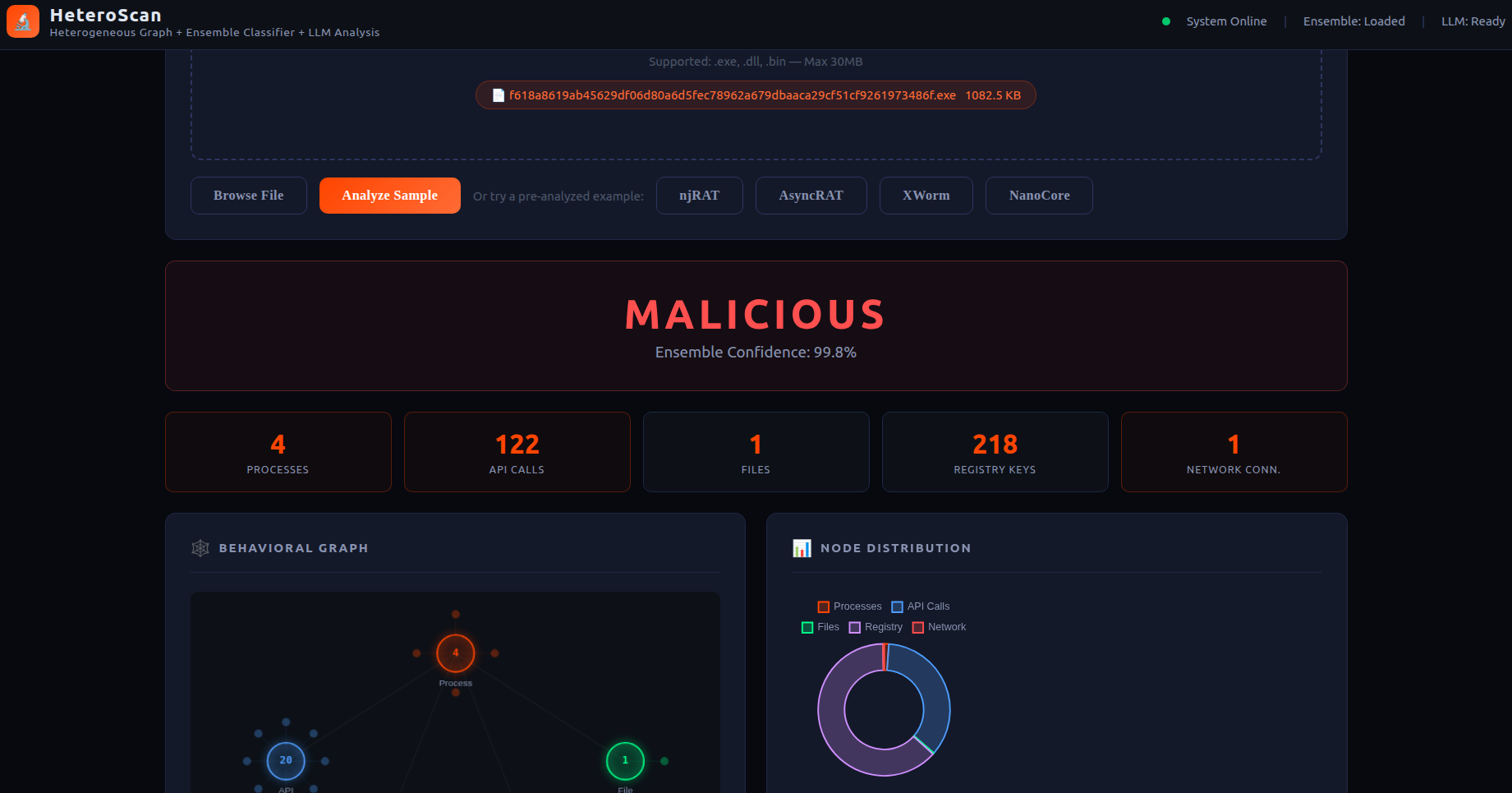}
\caption{HeteroScan analysis output for a NanoCore sample
showing the MALICIOUS verdict with 99.8\% ensemble
confidence, behavioral graph visualization, graph statistics,
and node distribution across process, API, file, registry,
and network entity types.}
\label{fig:heteroscan}
\end{figure*}
Registry-heavy families including BlackMatter (5,036),
QuasarRAT (1,190), WSHRAT (1,136), and DCRat (1,025)
perform extensive enumeration consistent with persistence
establishment. Network-active families including Stealc (4
connections), NanoCore, and Bandook (3 each) maintain more
external connections, consistent with information-stealing
and C2 functions. Process-heavy families including
BlackMatter, DCRat, NanoCore, QuasarRAT, and Babadeda show
injection and masquerading patterns. XWorm is a notable
outlier with only 14 registry accesses and 36 API calls,
indicating residual sandbox awareness.

\subsection{LLM Explanation Evaluation}

To evaluate the LLM explanation layer, we assessed
LLM-generated reports for 20 samples across 20 distinct
families against five behavioral consistency checks:
registry activity mentioned when accesses exceeded 100,
network behavior referenced when connections were observed,
process activity noted when more than two processes spawned,
no file encryption claims when no files were written, and
verdict matching classifier output. The layer achieved an
average score of 4.4 out of 5 across 20 samples, with 12
of 20 (60\%) passing all checks. Partial cases corresponded
to samples with sparse behavioral footprints where limited
graph evidence constrained explanation depth. No sample
produced a verdict inconsistent with classifier output.

\subsection{Case Studies}

Fig.~\ref{fig:heteroscan} shows HeteroScan analyzing a real
NanoCore sample, classified as MALICIOUS with 99.8\%
ensemble confidence. Table~\ref{tab:cases} summarizes the
behavioral graph statistics for three case studies.

\begin{table}[]
\caption{Behavioral Graph Statistics for Case Study Samples}
\label{tab:cases}
\centering
\begin{tabular}{lrrrrrc}
\toprule
\textbf{Family} & \textbf{Proc} & \textbf{API} &
\textbf{File} & \textbf{Reg} & \textbf{Net} &
\textbf{Conf} \\
\midrule
njRAT    & 3 & 133 & 1 & 461   & 3 & 99.9\% \\
AsyncRAT & 4 & 117 & 1 & 1,264 & 1 & 97.3\%  \\
NanoCore & 4 & 122 & 1 & 218   & 1 & 99.8\% \\
\bottomrule
\end{tabular}
\end{table}

\textbf{njRAT:} The sample spawned \texttt{svchost.exe} and
\texttt{dllhost.exe} as child processes, a masquerading
technique mapped to T1036. It made 133 unique API calls and
accessed 461 registry keys concentrated on
\texttt{HKLM\textbackslash SOFTWARE\textbackslash
Microsoft\textbackslash NET Framework Setup\textbackslash
NDP}, a pattern characteristic of .NET-compiled malware
performing runtime environment discovery. A connection to
\texttt{155.103.71.146} on port 777 was identified as a C2
beacon. The LLM explanation identified process injection via
\texttt{NtProtectVirtualMemory} and \texttt{LdrLoadDll} as
the primary threat indicator and mapped the behavior to
T1055, T1036, T1012, and T1071.

\textbf{AsyncRAT:} Execution was initiated through
\texttt{wscript.exe} before spawning
\texttt{MVXXXKRGFVHLOAUU.exe} and \texttt{dllhost.exe}. The
behavioral graph contained 1,264 registry key accesses
concentrated on .NET Framework configuration keys.
Techniques identified include T1059.005, T1112, T1036, and
T1041.

\textbf{NanoCore:} The sample spawned four processes and
made 122 unique API calls, accessing 218 registry keys
concentrated on \texttt{HKLM\textbackslash SOFTWARE\textbackslash
Microsoft\textbackslash .NETFramework}, consistent with
.NET runtime environment discovery. The behavioral graph
reveals a single network connection and minimal file
activity, indicating in-memory execution. The node
distribution shown in Fig.~\ref{fig:heteroscan} confirms
that registry and API interactions dominate the behavioral
footprint. Techniques identified include T1012, T1036,
T1055, and T1071. All three samples performed extensive
.NET Framework registry enumeration and spawned child
processes under legitimate system executable names,
patterns visible in the graph structure without examining
individual API semantics.

\subsection{Limitations}
The classification relies on aggregate structural statistics
rather than semantic node features, limiting family-level
discrimination for samples with similar topologies. The
dataset is predominantly commodity malware from
MalwareBazaar; APT samples and non-Windows malware are
underrepresented. The stratified split at the sample level
may be optimistic for closely related variants from the same
family appearing in both training and test partitions. The
sandbox analysis window may be insufficient to capture
delayed-trigger behavior. Several families including
Phorpiex and Netwire produced no usable behavioral graphs
despite sandbox hardening, reflecting genuine behavioral
minimalism or deep anti-analysis techniques. Behavioral
differences between malicious and benign samples may partly
reflect collection source rather than maliciousness alone.
\section{Conclusion}
\label{sec:conclusion}
This paper examined what malware actually does after gaining
access to a system, analyzing samples from 37 families
through dynamic sandbox execution using CAPEv2 and
heterogeneous graph representation. The proposed pipeline
converts behavioral traces into graphs with five node types
and seven edge relationships, extracts 55 structural
features, and classifies samples using a voting ensemble of
Random Forest, Extra Trees, and XGBoost. On 2,516 behavioral
graphs, the ensemble achieves 94.44\% accuracy and MCC of
0.863, outperforming a flat count baseline by 7.4 MCC
points. A multiclass classifier distinguishes 24 families
with macro F1 of 0.582, and an LLM explanation layer
evaluated on 20 samples achieves an average behavioral
consistency score of 4.4 out of 5. The behavioral fingerprint
analysis is perhaps the most direct answer to the paper's
central question: families like BlackMatter and DCRat leave
heavy registry footprints consistent with persistence
establishment, while Stealc and NanoCore maintain active
external connections, and families like QuasarRAT show
process spawning patterns consistent with injection. These
differences are visible at the graph topology level without
examining individual API semantics. The full pipeline is
deployed as HeteroScan at
\url{https://ama-andam.github.io/heteroscan}. Future work includes incorporating semantic node features to
enable GNN-based classifiers to close the performance gap
with the ensemble, expanding the dataset to cover APT
samples and non-Windows platforms, and developing temporal
heterogeneous graph representations that preserve execution
ordering within behavioral traces.

\bibliographystyle{IEEEtran}
\bibliography{references}

\end{document}